\documentclass{article}
\usepackage{spconf,amsmath}
\usepackage{amssymb,latexsym, fancyhdr, epsfig,float}
\usepackage{graphicx,cite}

\usepackage{amssymb,latexsym, subfig}
\def\beq{\begin{equation}} 
\def\eeq{\end{equation}}
\def\beqa{\begin{eqnarray}} 
\def\eeqa{\end{eqnarray}}

\def\bear{\begin{array}} 
\def\eear{\end{array}}
\def\ben{\begin{enumerate}}
\def\een{\end{enumerate}} 
\def\bit{\begin{itemize}}
\def\eit{\end{itemize}} 
\def\bfi{\begin{figure}}
\def\efi{\end{figure}} 

\def\ltc{\left\{} 
\def\rtc{\right\}}
\def\lts{\left[} 
\def\rts{\right]}

\def\teq{\triangleq}

\makeatletter
\def\BState{\State\hskip-\ALG@thistlm}
\makeatother

\def\br{\mbox{$\mathbf{r}$}}

\def\bx{\mbox{$\mathbf{x}$}}

\def\bLambda{\mbox{$\mathbf{\Lambda}$}}

\def\bv{\mbox{$\mathbf{v}$}}

\def\bw{\mbox{$\mathbf{w}$}}

\def\nn{\nonumber}

\long\def\gobbleup#1{}

\def\bn{\mbox{$\mathbf{n}$}}

\def\bTheta{\mbox{$\boldsymbol{\Theta}$}}

\def \bPsi{\mbox{$\boldsymbol{\Psi}$}}

\title{On Acoustic Modeling for Broadband Beamforming}
\name{Amit Chhetri, Mohamed Mansour, Wontak Kim, Guangdong Pan}
\address{Amazon Inc., USA}
\begin{document}

\ninept
\maketitle
\begin{abstract}
In this work, we describe limitations of the free-field propagation model for designing broadband beamformers for microphone arrays on a rigid surface. Towards this goal, we describe a general framework for quantifying the microphone array performance in a general wave-field by directly solving the acoustic wave equation. The model utilizes Finite-Element-Method (FEM) for evaluating the response of the microphone array surface to background 3D planar and spherical waves. The effectiveness of the framework is established by designing and evaluating a representative broadband beamformer under realistic acoustic conditions. 
\end{abstract}
\vspace{-0.1cm}
\section{Introduction}
\label{sec:intro}

Broadband beamforming with microphone array is a key signal processing module in many consumer electronics products, e.g., smart phones and smart speakers \cite{chhetri2018multichannel, MA_book, benesty2008microphone}. The proliferation of microphone arrays due to decreasing hardware cost and superior speech enhancement performance, has made broadband beamforming a ubiquitous embedded technology, and its performance has a critical impact on the overall system. 
 
A key requirement for broadband beamforming  is to deliver consistent performance across several octaves of frequencies, e.g., $80$ Hz - $8$ KHz in the voiceband case. Speech enhancement  is typically the system objective in most microphone arrays systems, rather than mere signal detection, as in the narrowband case. This poses hardware and algorithmic challenges in the design of microphone arrays and the underlying beamforming procedure. Filter-and-Sum (F\&S) \cite{frost_FilterandSum} has been a standard approach for designing a broadband beamformer as an extension to a narrowband beamformer by \emph{stitching} frequency-domain coefficients that are computed using narrowband beamforming techniques. Several narrowband beamforming techniques, with different objectives and assumptions, become standard design techniques, e.g., Delay-and-Sum (D\&S) \cite{DS, MA_book}, Minimum-Variance-Distortionless-Response (MVDR) \cite{MVDR1, MVDR2}, Subspace methods \cite{benesty2008microphone}. In this work, we do not address a particular beamformer design algorithm. Rather, the emphasis is on the acoustic modeling, which is common among these techniques. Without loss of generality, MVDR-based F\&S beamformer with a robustness constraint is used as a case study for our analysis. 
At a given frequency and \emph{look-direction}, the two key design parameters in almost all beamforming algorithms are the steering vector and the spatial coherence matrix. 
Proper design of these two parameters is the subject of this work. 

In Far-Field models, the acoustic wave is usually approximated by plane-waves \cite{RoomAcousticsBook}, and the steering vector at the direction/frequency of a plane wave is defined as the observed acoustic pressure at the different microphones when the microphone array is impinged with the plane-wave. Near-Field steering vectors can similarly be approximated by acoustic spherical waves. The observed wave-field in the general case is the superposition of the incident wave-field and the scattered wave-field. A typical approximation of the steering vector is the free-field approximation, which assumes sound propagation in free-field (at the speed of sound in air), and only the incident wave-field is considered. This approximation is used almost universally in the microphone array literature because it yields closed-form formulae that simplify beamformer analysis.  The main issue of the free-field approximation is that it ignores the impact of the device surface on the observed acoustic pressure, i.e., the scattered wave-field. This impact, as will be shown, can significantly change the microphone array behavior at certain frequencies and angles. 

A possible remedy to this problem is to rely on anechoic lab measurements to quantify the device response to incident waves. However, this is a time-consuming and high-cost solution, and imperfect experimental settings could lead to noticeable modeling errors, especially in near-field cases. In this work, we describe a simulation-based approach for acoustic modeling 
of microphone array on rigid surface by solving the Helmholtz wave equation using Finite-Element-Method (FEM) with a background wave-field that matches the incident wave-field \cite{larsson2008partial}.  Prior works that studied the impact of scattered field on microphone arrays used spherical harmonic decomposition for specific form-factors (e.g. sphere, cylinder) \cite{zotkin2017incident, teutsch2007modal, rafaely2015fundamentals} (and references, therein). However, these methods are restrictive in the choice of device form-factors (e.g. do not include modern smart-speaker form-factors) and beamforming techniques. In comparison, the FEM method proposed in this paper provides three notable contributions: (i) a methodology to compute the steering vector for microphone arrays mounted on solid hard surfaces without the need for expensive anechoic chamber measurements, (ii) ability to design any type of beamformer that  relies on steering vectors; these include MVDR beamformer \cite{MA_book}, linearly constrained minimum variance (LCMV) beamformer \cite{VanTrees2002,mabande2009design}, and polynomial beamformer \cite{mabande2010design}, and (iii) extension of the proposed method to generic device form-factors that are used for smart speakers. 

The following notations are used throughout the paper. A bold lower-case letter denotes a column vector, while a bold upper-case letter denotes a matrix. ${\bf{A}}^T$ and ${\bf{A}}^H$ denote the transpose and conjugate transpose, respectively, of $\bf{A}$, and ${\bf{A}}_{m,n}$ is the matrix entry at position $(m,n)$. $\Theta \triangleq \left(\theta, \phi \right)^T$ denotes the polar and azimuth angles, respectively, in a spherical coordinate system. $M$ always refers to the number of microphones.  $\bPsi (\omega)$ denotes the noise coherence matrix, of size $M\times M$, at frequency $\omega$ (the dependency on $\omega$ is dropped whenever it is clear from the context).  Additional notations are introduced when needed.

\section{Background}
\label{sec:background}
\subsection{Wave Equation}
The acoustic wave equation \cite{acousticsbook} is the governing equation for the propagation of sound waves at equilibrium in elastic fluids, e.g., air. The homogenous wave equation has the  form
\begin{equation}
\nabla^2 \bar{p} - \frac{1}{c^2} \frac{\partial^2 \bar{p} }{\partial t^2} = 0
\end{equation}
where $\bar{p}(t)$ is the acoustic pressure, and $c$ is the speed of sound in the medium. In this work, we consider only the practical case of homogenous fluid with no viscosity. 

In practice, the wave equation is usually solved in the frequency domain using the Helmholtz equation to find $p(\omega)$:
\begin{equation}
\nabla^2 p + k^2 p = 0  \label{eq:Helmholtz}
\end{equation}
where $k\triangleq\omega/c$ is the wave number. At steady state, the time-domain and frequency-domain solutions are Fourier pairs \cite{FourierAcoustics}. In our modeling, we work only with the homogenous Helmholtz equation under various boundary conditions. The boundary conditions are determined by the geometry and the acoustic impedance of the different boundaries. We assume the device has a rigid surface, therefore, it is modeled as a \emph{sound hard} boundary.
\vspace{-0.1cm}
\subsection{Beamforming Strategies} \label{strategies}
Beamforming is a microphone-array signal processing technique that allows emphasizing the user's speech from a desirable look-direction (LD) while suppressing interferences from other directions. Here, we process microphone elements such that the signals arriving from look-direction are combined in-phase, while signals arriving from other directions are combined out-of-phase. 
Denote the position of the $m$-th microphone by $\br_m$, and the signal acquired at the $m$-th microphone for frequency $\omega$ by  $x(w,\br_m)$. Then, the signal acquired by the microphone array can then be expressed as:
\beq \label{eq:xdef}
\bx(\omega,\br) = \lts x(\omega,\br_1)  \hspace{2mm} x(\omega,\br_2)  \hspace{2mm} \hdots \hspace{2mm} x(\omega,\br_M) \rts^T .
\eeq
Denoting the spectrum of the desired source signal by $s(\omega)$ and the ambient noise captured by the microphone array as $\bn(\omega)$, we can express $\bx(\omega,\br)$ as:
\beq \label{eq:XExpand}
\bx(\omega,\br) = \bv(\omega,\bTheta) s(\omega) + \bn(\omega),
\eeq
where, $\bv(\omega,\Theta) \teq \lts v_1(\omega,\Theta)  \hspace{2mm} v_2(\omega,\Theta) \hspace{2mm} \hdots \hspace{2mm} v_M(\omega,\Theta) \rts$ is the frequency and angle-dependent steering vector. 

The beamformer design involves computation of complex-valued weights for each frequency and microphone denoted by $\bw (\omega)  \teq \lts w_1(\omega)  \hspace{2mm}  w_2(\omega) \hspace{2mm}  \hdots \hspace{2mm}  w_M(\omega) \rts^T$, which are then applied to $\bx(\omega,\br)$ to obtain the beamformer output $y(\omega)$:
\beq \label{eq:bfOut}
y(\omega) = \bw^H (\omega) \bx(\omega,\br).
\eeq

We are interested in using FEM modeling for the design of F\&S beamformers that can be expressed as a constrained optimization problem, and the solution to which provides the optimal beamformer filters. This covers various F\&S beamformers like MVDR, maximum SNR, and LCMV beamformers \cite{VanTrees2002}. In this work, we use, without loss of generality,  the MVDR beamformer with a robustness constraint to present our analysis.

\subsection{Beamforming Metrics} \label{metrics}
We use three metrics to assess the performance: array gain (AG), white noise gain (WNG)  \cite{MA_book}, and microphone array channel capacity (MACC) \cite{mansour2018information}. 
The AG metric is defined as the improvement in signal-to-noise-ratio (SNR) offered by the beamformer:
$\text{AG}(\omega) \teq \frac{SNR_{out}(\omega)}{SNR_{in}(\omega)}$.
After some algebraic manipulations, one can show that \cite{MA_book}:
\beq \label{eq:AG}
\text{AG}(\omega, \Theta_{LD})  = \frac{|\bw^H \bv(\omega,\Theta_{LD})|^2}{\bw^H \bPsi(\omega) \bw},
\eeq
where $\Theta_{LD}$ denotes the look-direction, and $\bPsi \teq \bLambda/\beta$ is the normalized noise correlation matrix with
\beq \label{eq:LambdaNN}
\bLambda_{m,q} = \int_{0}^{\pi} \int_{0}^{2\pi} \bv_m(\omega,\Theta) \bv_q^{*}(\omega,\Theta) \sigma_{N}^{2} (\omega,\Theta) \sin(\theta) \ d\theta \ d\phi ,
\eeq
where $\sigma_{N}^{2} (\omega,\Theta)$ denotes the distribution of noise power as a function of $\omega$ and $\Theta$, and
\beq \label{eq:phiN}
\beta = \int_{0}^{\pi} \int_{0}^{2\pi} \sigma_{N}^{2} (\omega,\Theta) \sin(\theta) \ d\theta \ d\phi .
\eeq
The WNG metric is the SNR improvement provided by the beamformer when the noise components at the microphones are statistically independent \cite{MA_book}:
\beq \label{eq:WNG}
\text{WNG}(\omega, \Theta_{LD}) = \frac{|\bw^H \bv(\omega,\Theta_{LD})|^2}{\bw^H \bw}.
\eeq
The MACC metric \cite{mansour2018information} aims at providing a characterization of the microphone array that is independent of the beamformer realization. It is analogous to MIMO channel capacity in wireless communication. If the source location is known, then the MACC is defined as 
\begin{equation}
\text{MACC}(\omega, \Theta_{LD}) \triangleq \log \left(1+ P\|\mathbf{S}^{-\frac{1}{2}} \mathbf{U}^H \mathbf{v}(\omega, \Theta_{LD})\|^2\right)
\end{equation} 
where $\mathbf{USU}^H$ is the singular value decomposition of $\bPsi$, and $P$ is the input power.

\section{Acoustic Modeling}
\label{sec:typestyle}
\subsection{Acoustic Plane-Waves}
Acoustic plane waves constitute a powerful tool for analyzing the wave equation, and it provides a good approximation of the wavefield emanating from a far-field point source \cite{teutsch2007modal}. The acoustic pressure of a plane-wave with vector wave number $\bf{k}$ is defined at a point  $\bf{r}$ in the 3D space  as:
\begin{equation}
p({\bf{k}}) \triangleq p_0 e^{-j {\bf{k}}^T{\bf{r}}}   \label{eq:pw_def}
\end{equation} 
This is a solution of the inhomogeneous Helmholtz equation with a far point source, where $\|{\bf{k}}\| = k$ (note that, for a given $k$ in the Helmholtz equation, there are two degrees of freedom in choosing $\bf{k}$). Further, a general solution to the homogenous Helmholtz equation can be approximated by a linear superposition of plane waves of different angles \cite{FourierAcoustics, pwd1, pwd2, pwd3}. These properties render acoustic plane-waves a key tool in designing far-field beamforming for microphone arrays, where the microphone array response to each plane wave provides a sufficient set for the beamformer design.

The \emph{total} wavefield at each microphone of the microphone array when an incident plane-wave $p_i({\bf{k}})$ impinges on the device has the general form:
\vspace{-0.1cm}
\begin{equation}
p_t = p_i + p_s  \label{eq:pt_df}
\vspace{-0.1cm} 
\end{equation}
where $p_t$ and $p_s$ refer to the total and scattered wavefield respectively. The total wavefield, $p_t$, at each microphone is computed by inserting \eqref{eq:pt_df} in the Helmholtz equation \eqref{eq:Helmholtz} and solving for $p_s$ with appropriate boundary conditions. The details of this modeling are described in section \ref{sec:FEM_modeling}. It is evident from \eqref{eq:pw_def} that an incident plane-wave does not have magnitude information, and it is fully parameterized by its phase. This is not true for the scattered wavefield, $p_s$ which represents the reflections/diffractions due to the rigid device surface. This magnitude information in $p_s$ is critical in resolving phase ambiguity due to microphone array geometry. 

If the microphone array is composed of discrete microphones in space, and the area of each microphone is much smaller than the wavelength, then a reasonable approximation is to set $p_s = 0$ in \eqref{eq:pt_df}. This is referred to as \emph{free-field} approximation. In this case, the total wavefield, $p_t$, is fully determined by the wavenumber $\bf{k}$ in \eqref{eq:pw_def}, and the $(x,y,z)$ coordinates of each microphone. It is obvious that, free-field approximation is not accurate if the microphone array is on a rigid surface. Nevertheless, this approximation has been utilized almost universally in the literature for acoustic modeling in beamformer design.  In the following section, we show that the free-field approximation does not provide a good approximation of the total field under important practical cases.

\vspace{-0.2cm}
\subsection{FEM Modeling}
\label{sec:FEM_modeling}

The modeling objective  is to compute the total sound field in \eqref{eq:pt_df} at each microphone when the device is impinged by a plane wave. This resembles physical measurement in anechoic room with a distant point  source. FEM is one of the standard approaches for solving the Helmholtz equation numerically. In our case, we need to solve the Helmholtz equation for the total wavefield at all frequencies of interest  with a \emph{background} plane wave. The device surface is modeled as sound hard boundary. The microphone is modeled as a point receiver on the surface if the microphone surface area is much smaller than the wavelength, otherwise, its response is computed as the integral of the acoustic pressure over its area. To have a true background plane-wave, the external boundary should be open and non-reflecting. In our model, the device is enclosed by a closed boundary, e.g., a cylinder or a spherical surface. To mimic open-ended boundary, there are two choices:  (i) Matched boundary whose impedance is matched to the air impedance at the frequency of interest, (ii) Perfectly matched layer,  which defines a special absorbing domain that eliminates reflection and refractions in the internal domain that encloses the device \cite{berenger1994perfectly}.  The merits of each approach is beyond the scope of this paper. The FEM solves for $p_t$ in \eqref{eq:pt_df}, which is equivalent to solving for only the scattered field, $p_s$, after inserting background plane wave model \eqref{eq:pw_def} in the Helmholtz equation. The acoustics module of COMSOL multiphysics package \cite{COMSOL} is used for this FEM numerical solution, and the simulation is rigorously validated  with exact and measured results on different form-factors. For example, in Fig. \ref{fig:comsol}, we show the total pressure field of two microphones on a spherical surface with analytical and simulated solution. Both amplitude and phase responses match excellently with the analytical solution \cite{ Bowman_book}.  Further, in Fig. \ref{fig:comsol2}, we show an example of simulated and measured acoustic pressure of a rectangular microphone array mounted on a slanted cube. In the plot, we show the inter-channel response, i.e., $\{H_i(\omega)/H_r(\omega)\}_{i\neq r}$, where $r$ is a reference microphone. The phase difference between  simulated and measured responses is linear, which is expected when the positions of the device in both cases are not perfectly aligned.  For more comparisons between simulated/theoretical and measured acoustic pressure responses, one may refer to \cite{wiener1949diffraction,spence1948diffraction}.
\bfi[t] 
	\centering
	\includegraphics[width=0.4\textwidth, trim=5mm 5mm 5mm 5mm,clip]{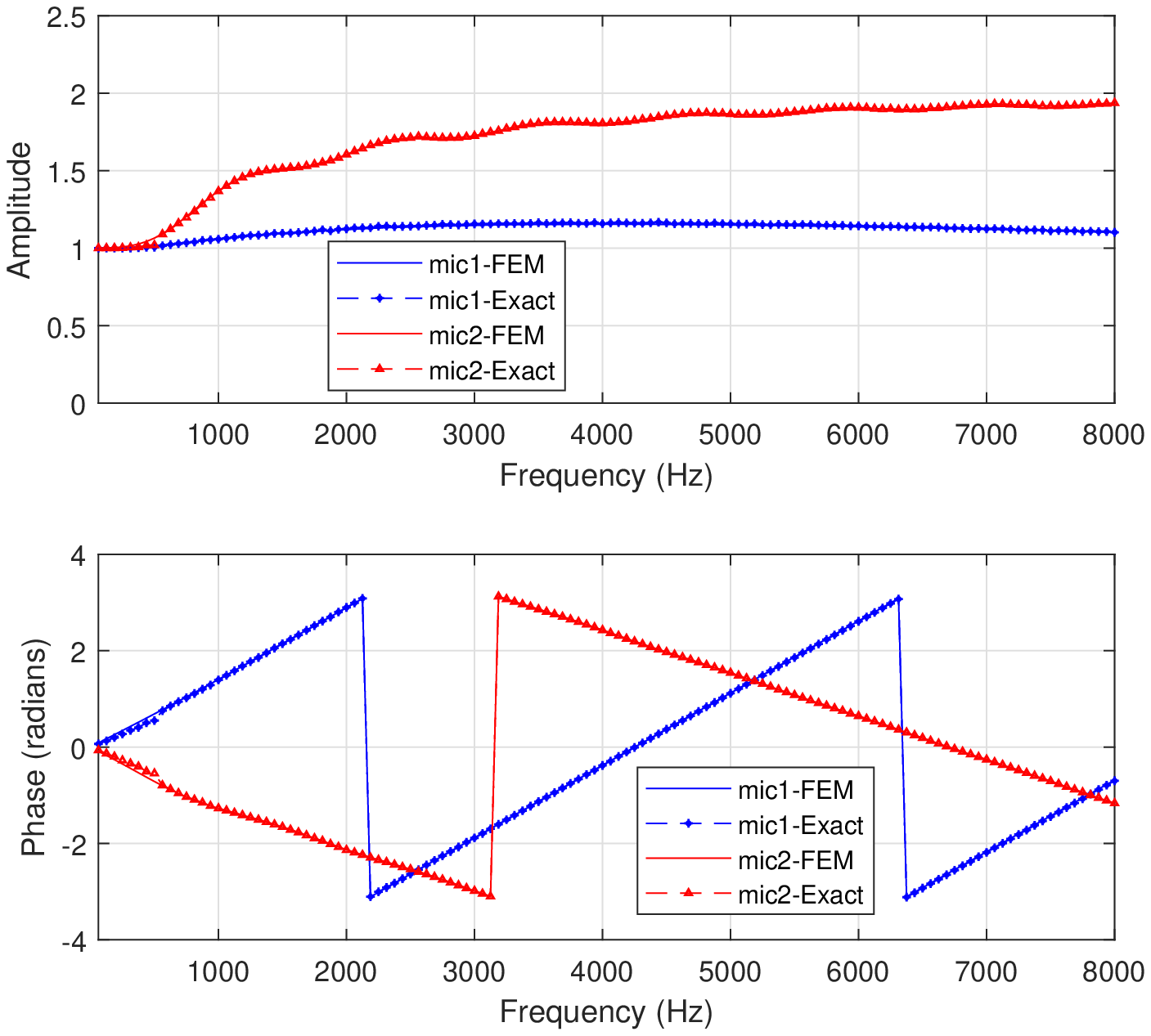}
	\caption{\footnotesize{Comparison of FEM and analytical solutions for spherical surface of radius 5 cm. Top: magnitude response, bottom: phase response. Mic 2 is in the middle of sphere facing the incident wave; Mic 1 is facing away from it.} } 
	\label{fig:comsol}
\efi
\bfi[htp] 
	\centering
	\includegraphics[width=0.45\textwidth, trim=5mm 1mm 5mm 7.8mm,clip]{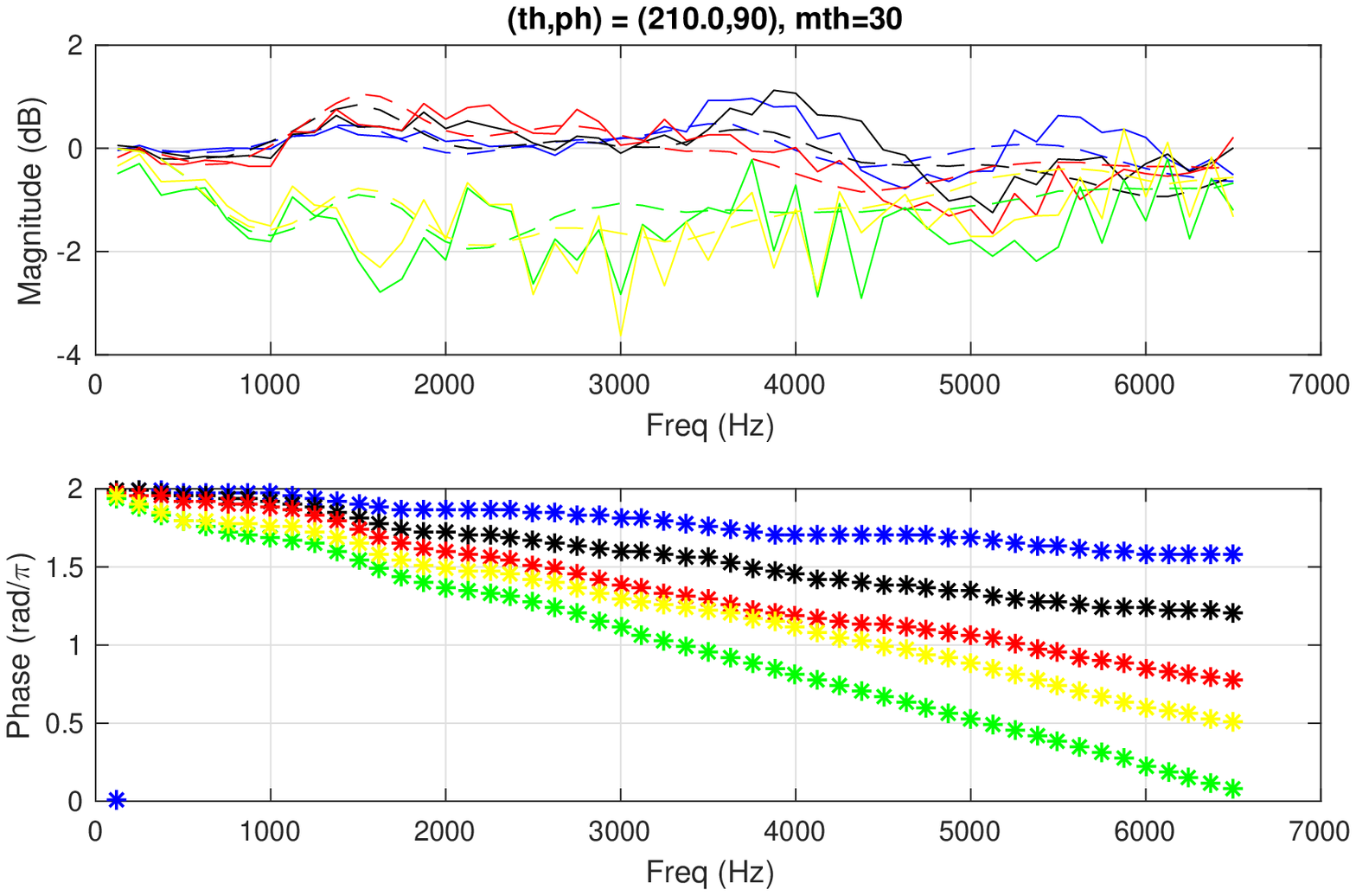}
	\caption{\footnotesize{Normalized acoustic pressure of a rectangular microphone array on a slanted cube for simulated and measured cases for a background plane wave at direction $(\theta,\phi) =(90^\circ, 210^\circ)$ (simulated response is in dotted lines, measured response in solid lines, with different colors for different microphones). Top: magnitude, bottom: phase difference} } 
	\label{fig:comsol2}
\vspace{-0.25cm}
\efi

Note that, the above procedure is not limited to plane-wave as we only need to specify the background pressure field, which could be, for example, spherical wave for near-field modeling. The procedure is repeated for a grid of frequency and incident angles to build a dictionary of total pressure  $\{p_t(\omega, \theta,\phi)\}_{\omega, \theta, \phi}$ that is used in subsequent analysis.

\vspace{-0.25cm}
\section{Analysis of Free-Space Beamforming}
\label{sec:analysis}
To illustrate the benefits of FEM modeling, we use the MVDR beamformer with a robustness constraint,  formulated as a constrained convex optimization problem \cite{mabande2009design}:
\vspace{-0.1cm}
\beqa  \label{eq:MVDR}
\widehat\bw = \arg \min_{\bw} & & \bw^H \bPsi \ \bw \nn \\
 \text{such that} & & \bw^H \bv(\omega, \theta_{LD}) =1, \nn \\
 & & \frac{|\bw^H \bv(\omega,\Theta_{LD})|^2}{\bw^H \bw}  \geq \gamma ,
 \eeqa
 where the first constraint is called the distortionless constraint \cite{ MA_book}, and  the second constraint is the WNG constraint, which imposes robustness in the beamformer design that can be controlled through $\gamma$ \cite{mabande2009design}. Further, the WNG constraint enables a more fair comparison between the total and free-field beamformer designs because the WNG is bounded in both cases. Without loss of generality, we assume a spherically diffuse noise field. The optimization problem in \eqref{eq:MVDR} is solved using a convex optimization solver to obtain the beamformer weights $\widehat\bw$. Note that the proposed FEM-model based method can be similarly extended to other beamformer designs like the MVDR \cite{MA_book}, LCMV \cite{VanTrees2002}, and polynomial beamformer \cite{ mabande2010design}.

\subsection{Analysis Methodology}
\label{sec:anaMethod}
 The MVDR solution can be obtained from \eqref{eq:MVDR} by using $\bv_i$ and $\bv_t$ as steering vectors for free-field (FF) and total-field (TF), respectively. To compute $\bPsi$, we use analytical method for canonical device shapes, such as finite cylinder and sphere \cite{Bowman_book}. For a general device shape, the FEM tool is used to simulate the steering vectors for a uniform grid of azimuth and polar angles. Then, $\bPsi$ is numerically computed from \eqref{eq:LambdaNN} and \eqref{eq:phiN}, with $\sigma_N^2(\omega,\Theta)=1$ for the spherically diffuse noise field.
 
 \bfi[t] 
	\centering
	\includegraphics[width=9cm, height=3.25cm, trim=15mm 5mm 5mm 30mm,clip]{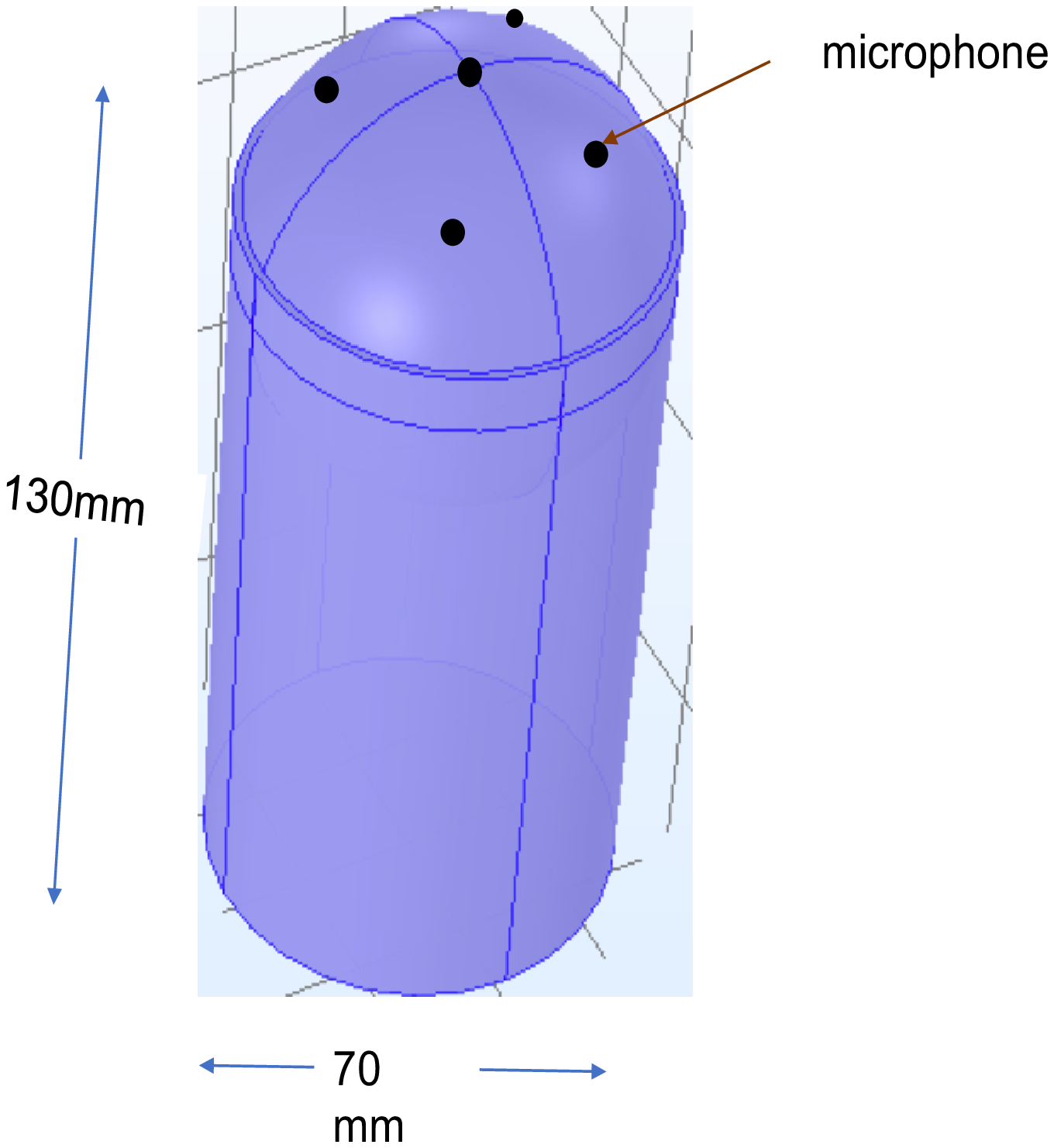}
	\vspace*{-0.5cm}
	\caption{Simulation setup for FEM-based beamforming. The form-factor is a combination of a cylindrical bottom and a top surface with a spherically-curved shape. }
	\label{fig:arraySetup}
\efi

We now compare the performance of MVDR beamformer designed using FF and TF assumptions. For our study, we use the setup in  Fig. \ref{fig:arraySetup}, which has $5$ microphones on the top of a cylinder of height 130 mm and diameter of 70 mm; the top surface of the cylinder has a spherically-curved shape. This surface does not have a closed-form solution for the Helmholtz equation, which necessitates the use of the proposed FEM method. The origin of the coordinate system coincides with the center microphone with $z$ axis pointing upwards, and the $x$-$y$ plane parallel to the bottom face of the cylinder. The coordinates of the microphones are: $(x,y,z)= \ltc (r_o,0,z_o), (0,r_o,z_o), (-r_o,0,z_o),(0,-r_o,z_o), (0,0,0) \rtc$, where $r_o=30$ mm and $z_o=-3$ mm. Lastly, we set $\gamma=-25$ dB. 
\vspace{-0.25cm}

\subsection {Results} \label{sec:results}
\begin{figure}[h]%
    \centering
    \subfloat[$(\theta,\phi) = (90^\circ, 0^\circ)$]{{\includegraphics[width=4.295cm, height=3.7cm, trim=11mm 0mm 11mm 7mm,clip]{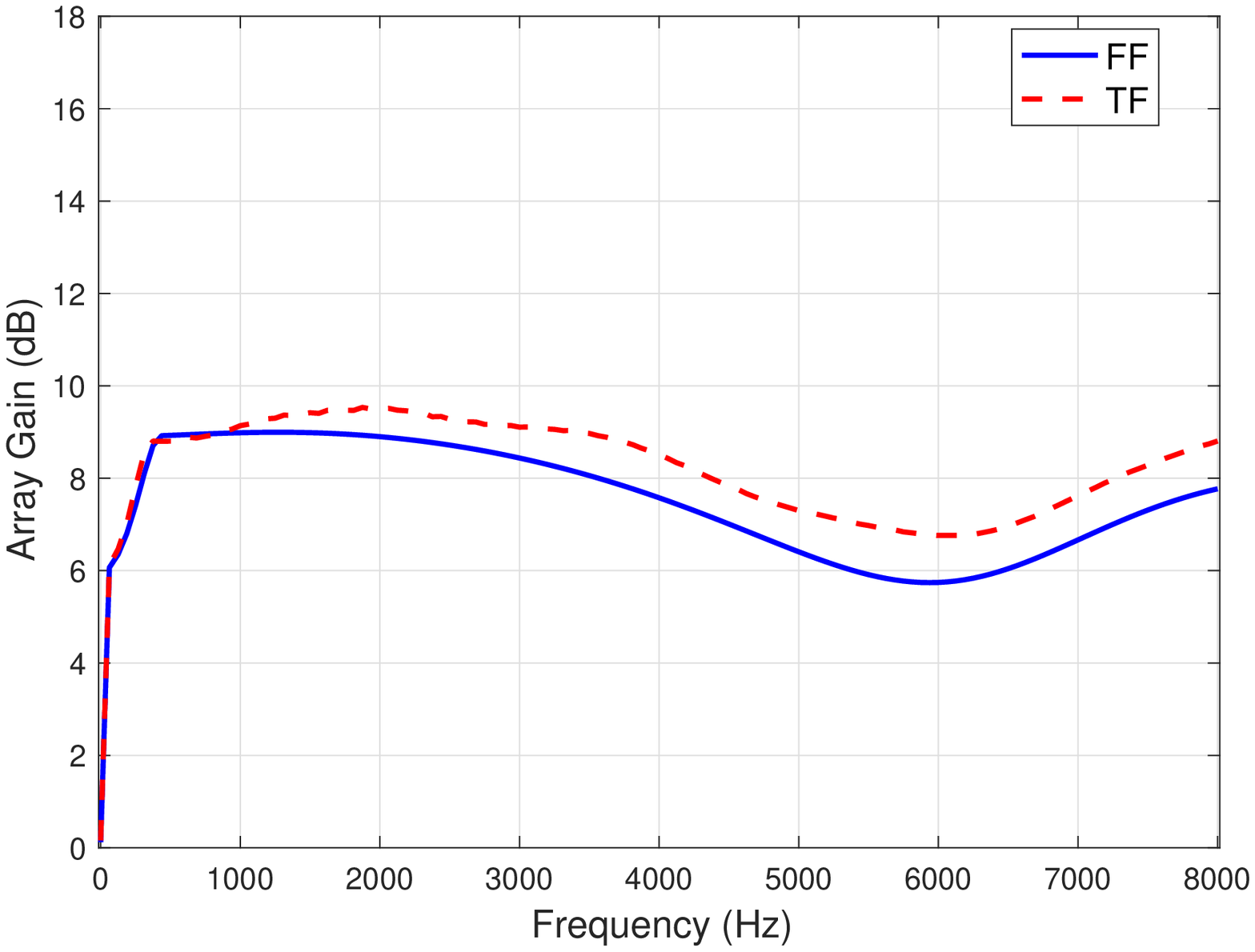} }}
    \subfloat[$(\theta,\phi) = (30^\circ, 0^\circ)$]{{\includegraphics[width=4.295cm, height=3.7cm,trim=11mm 0mm 11mm 7mm,clip]{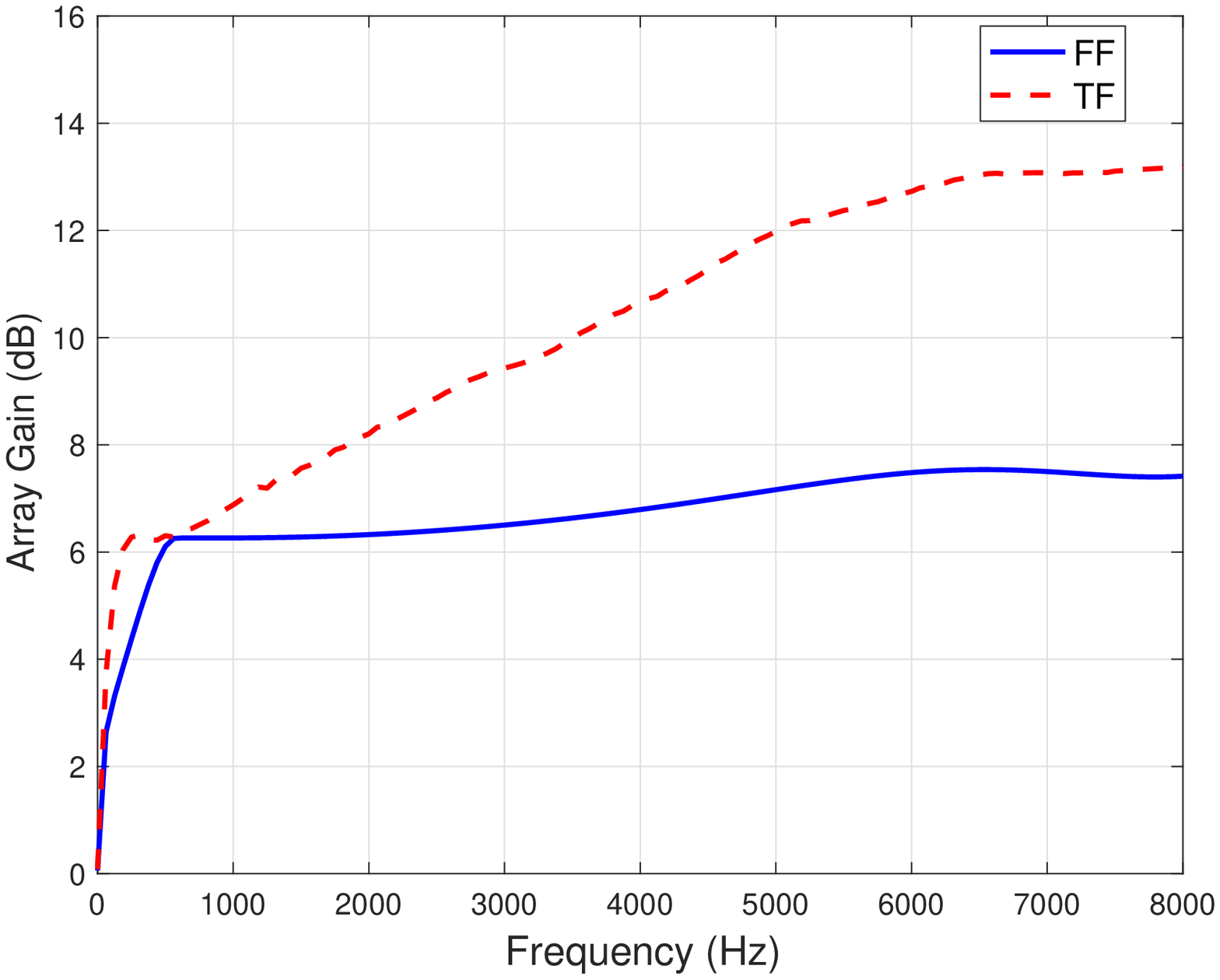} }}
	\vspace*{-0.15cm}
    \caption{AG performance showing the difference between FF and TF. Note how the array gain is higher, especially at high frequencies. }
    \label{fig:AG}%
	\vspace*{-0.3cm}
\end{figure}

\begin{figure}[h]%
    \centering
    \subfloat[$(\theta,\phi) = (90^\circ, 0^\circ)$]{{\includegraphics[width=4.295cm, height=3.7cm, trim=11mm 0mm 11mm 7mm,clip]{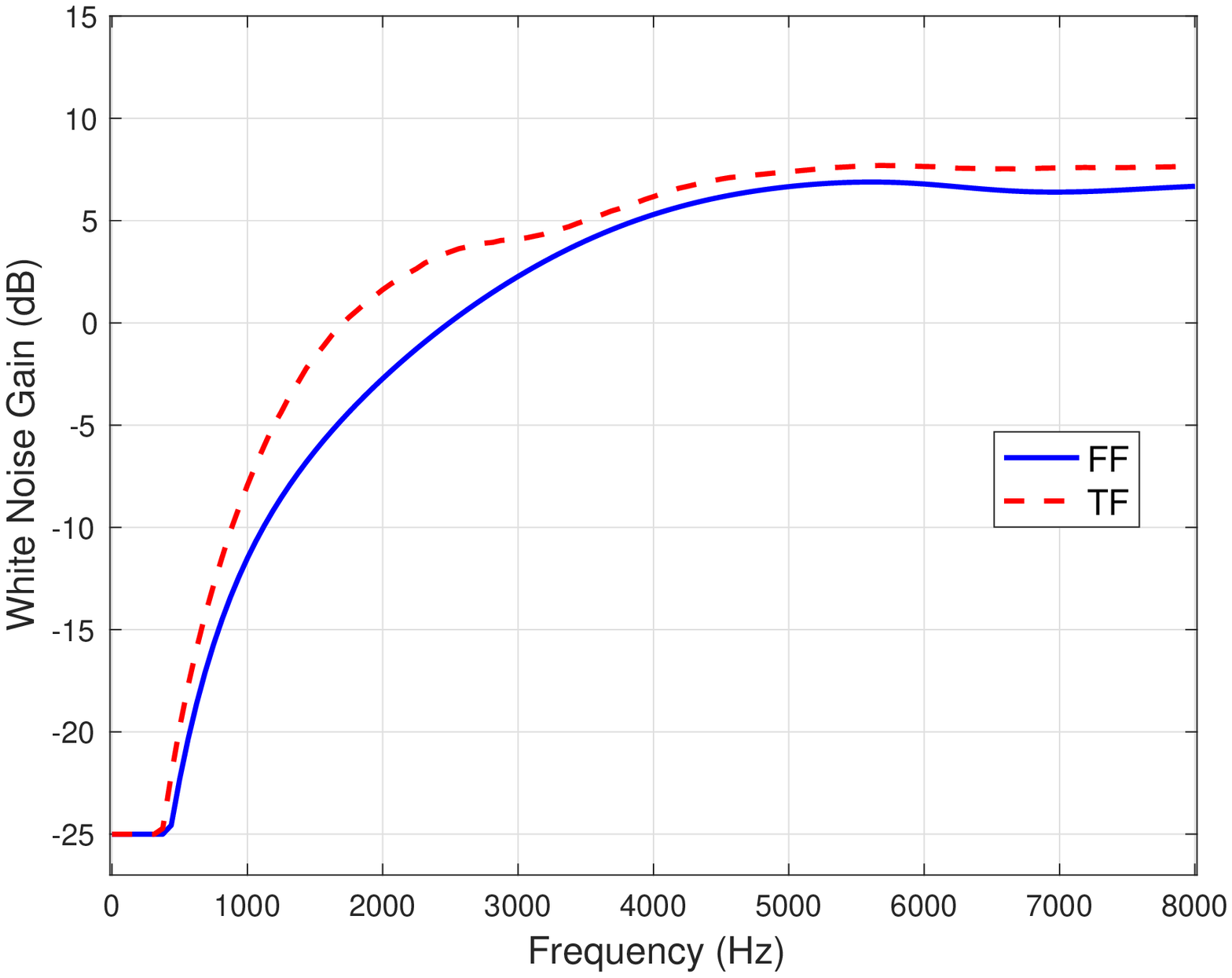} }}
    \subfloat[$(\theta,\phi) = (30^\circ, 0^\circ)$]{{\includegraphics[width=4.295cm, height=3.7cm,trim=11mm 0mm 11mm 7mm,clip]{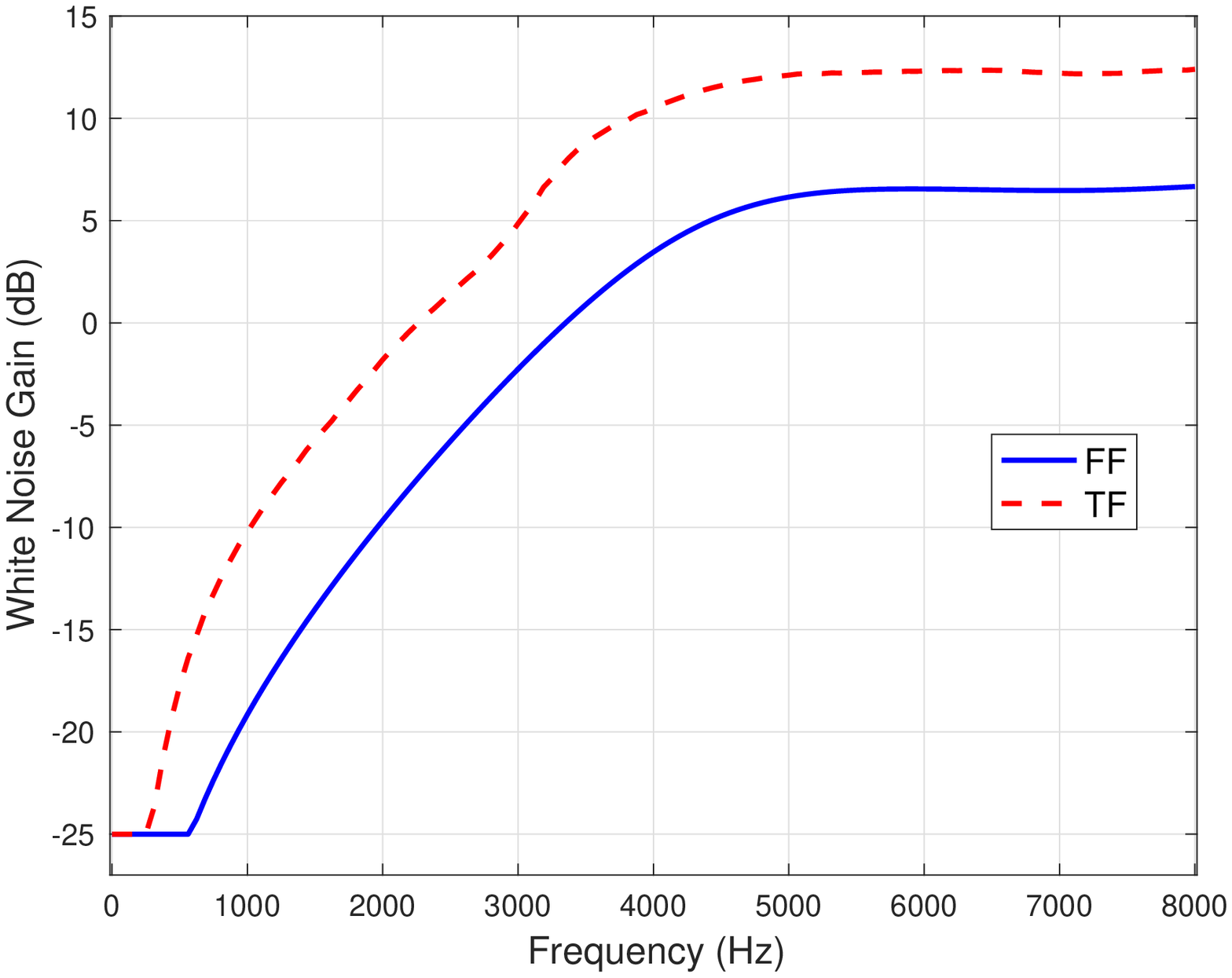} }}
	\vspace*{-0.15cm}
    \caption{WNG performance. Note that even with the higher array gain from Fig. \ref{fig:AG}, the WNG is better for the TF configuration.}
    \label{fig:WNG}%
	\vspace*{-0.3cm}
\end{figure}

\begin{figure}[h]%
    \centering
    \subfloat[$(\theta,\phi) = (90^\circ, 0^\circ)$]{{\includegraphics[width=4.295cm, height=3.7cm, trim=11mm 0mm 11mm 7mm,clip]{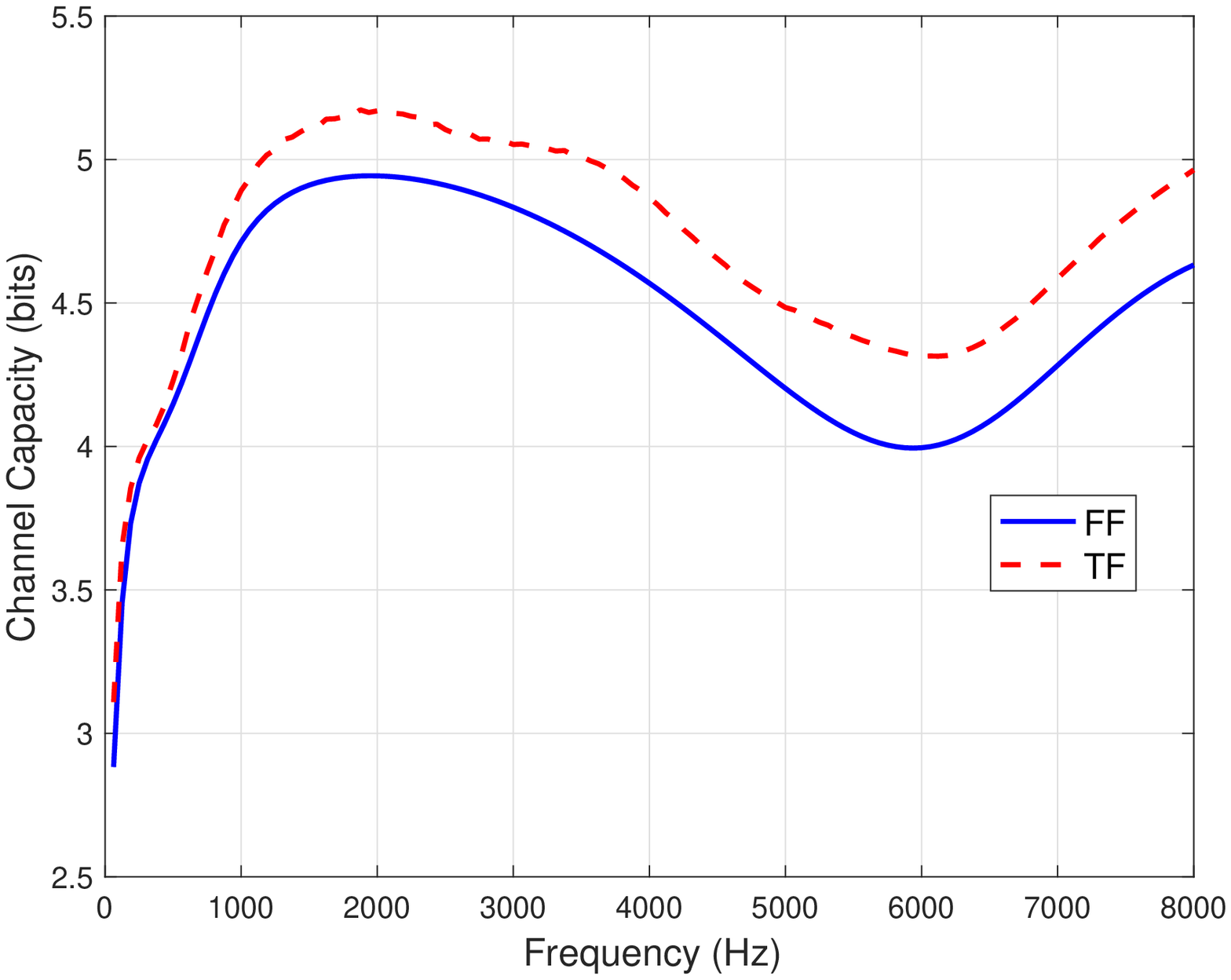} }}
    \subfloat[$(\theta,\phi) = (30^\circ, 0^\circ)$]{{\includegraphics[width=4.295cm, height=3.7cm,trim=11mm 0mm 11mm 7mm,clip]{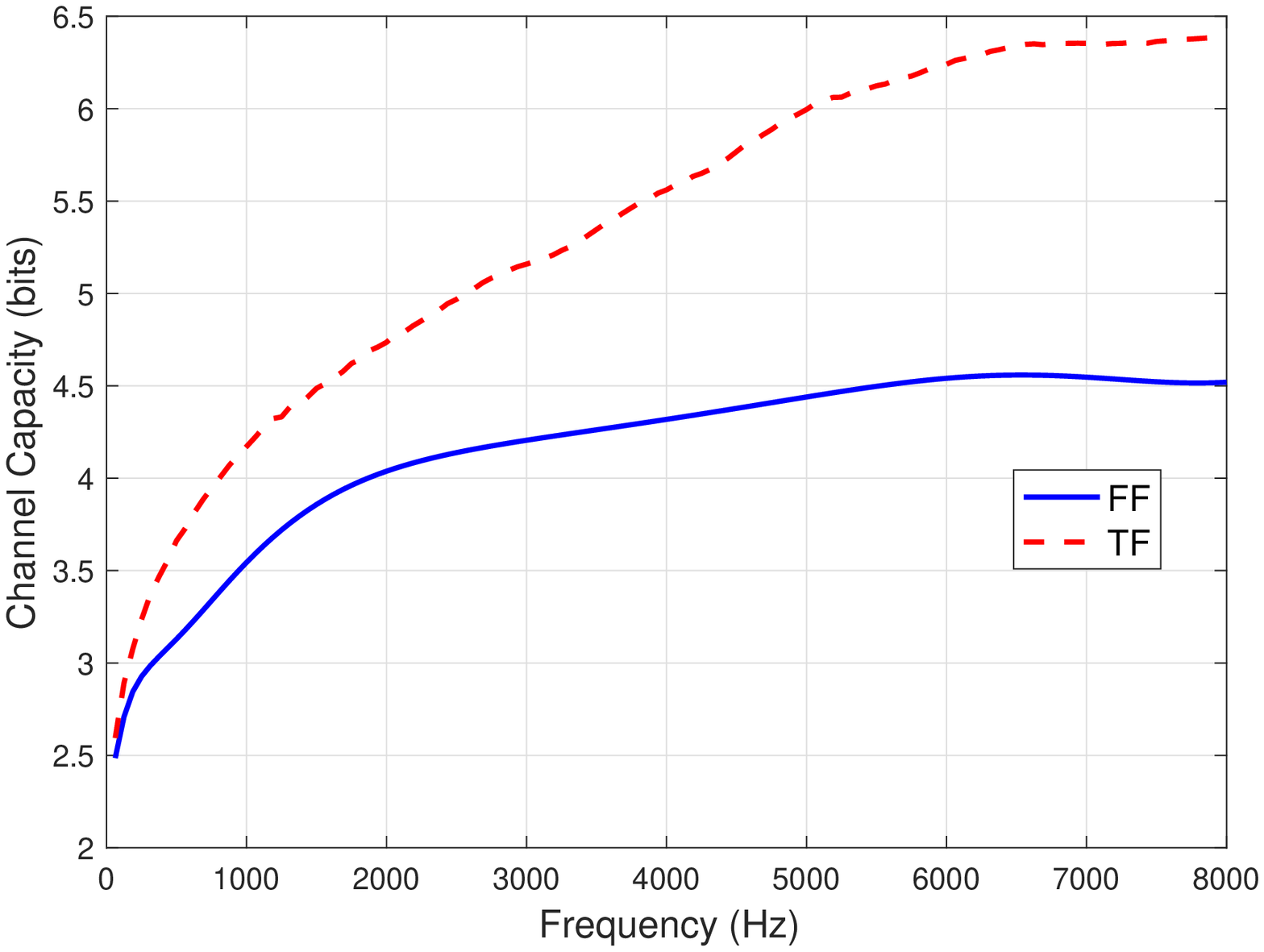} }}
    	\vspace*{-0.15cm}
\caption{MACC performance, which corresponds well with the AG performance in Fig. \ref{fig:AG}.}
    \label{fig:MACC}%
	\vspace*{-0.3cm}
\end{figure}

We evaluate the microphone array metrics under Free Field (FF) and Total Field (TF) setups for the array in Fig. \ref{fig:arraySetup} at two arrival angles: $(\theta,\phi) = (90^\circ, 0^\circ)$ and $(30^\circ, 0^\circ)$. The results are summarized in Figs. \ref{fig:AG}-\ref{fig:MACC} for the three performance metrics. 

 
 At $\theta = 90^\circ$, i.e., $x$-$y$ plane, the TF case is slightly better for the AG and MACC, but the WNG performance for the TF case is better than the FF case. This is explained by noting that the steering vectors in the TF case have variations in both phase and amplitude (over microphones) in comparison to the FF case, which only has phase variations. The amplitude variations increase the spatial diversity for the TF case, which can be used to improve the spatial directivity of the beamformer.
At $\theta = 30^\circ$, the TF case has WNG performance better than the FF over the full frequency range; the AG for TF case is noticeably better than FF case for all frequencies, and significantly better for frequencies beyond 2 kHz.  Note that the WNG curves are lower-bounded by $-25$ dB, because of the WNG constraint specified in (\ref{eq:MVDR}). Note also that in all cases, the MACC for the TF case is noticeably better than the FF case, because the FF case ignores the magnitude information, which provides invaluable characterization of the look direction. 

\begin{figure}[h]%
    \centering
    \subfloat[$(\theta,\phi) = (90^\circ, 0^\circ)$]{{\includegraphics[width=4.295cm, height=3.7cm, trim=10mm 0mm 11mm 7mm,clip]{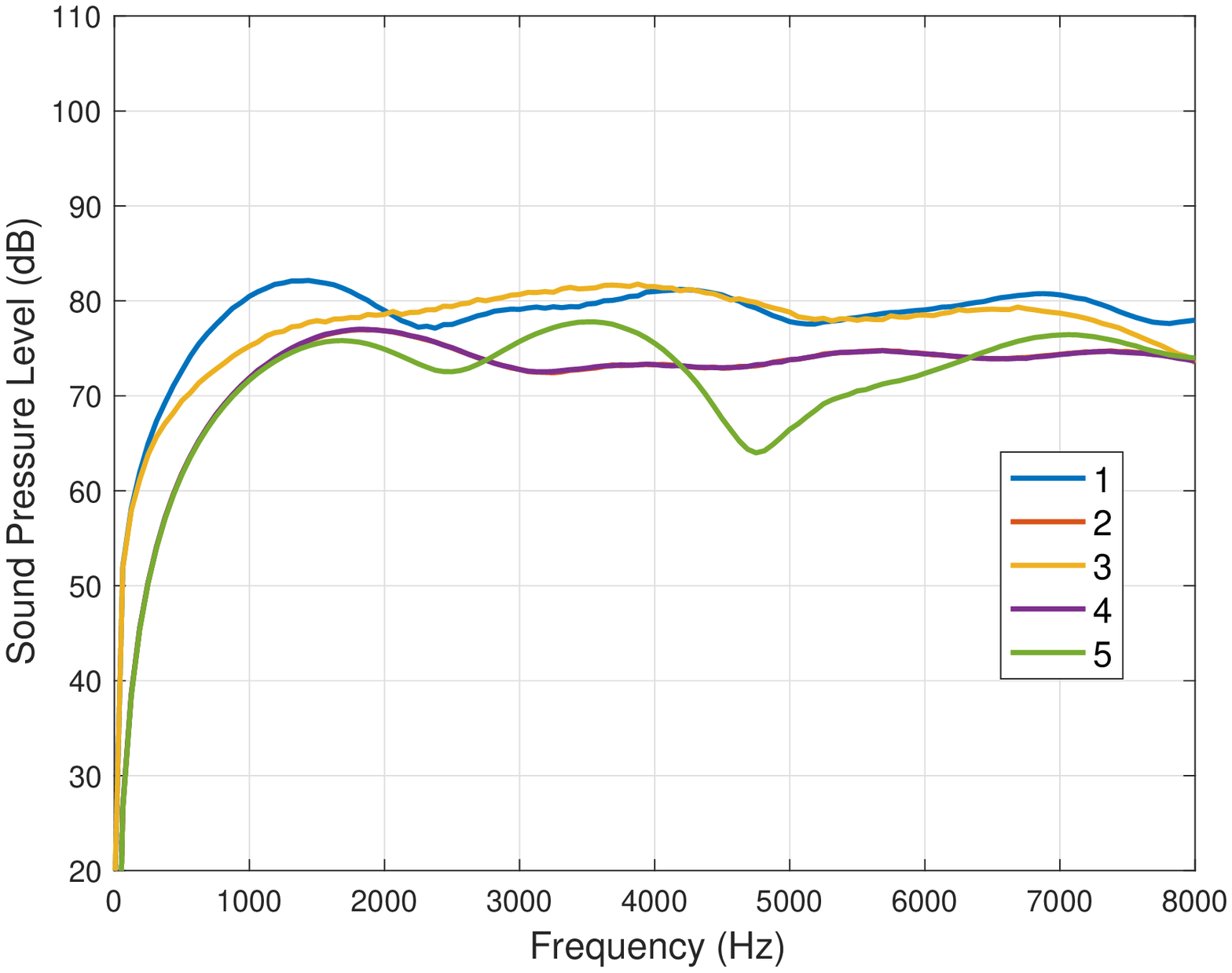} }}%
    \subfloat[$(\theta,\phi) = (30^\circ, 0^\circ)$]{{\includegraphics[width=4.295cm, height=3.7cm,trim=10mm 0mm 11mm 7mm,clip]{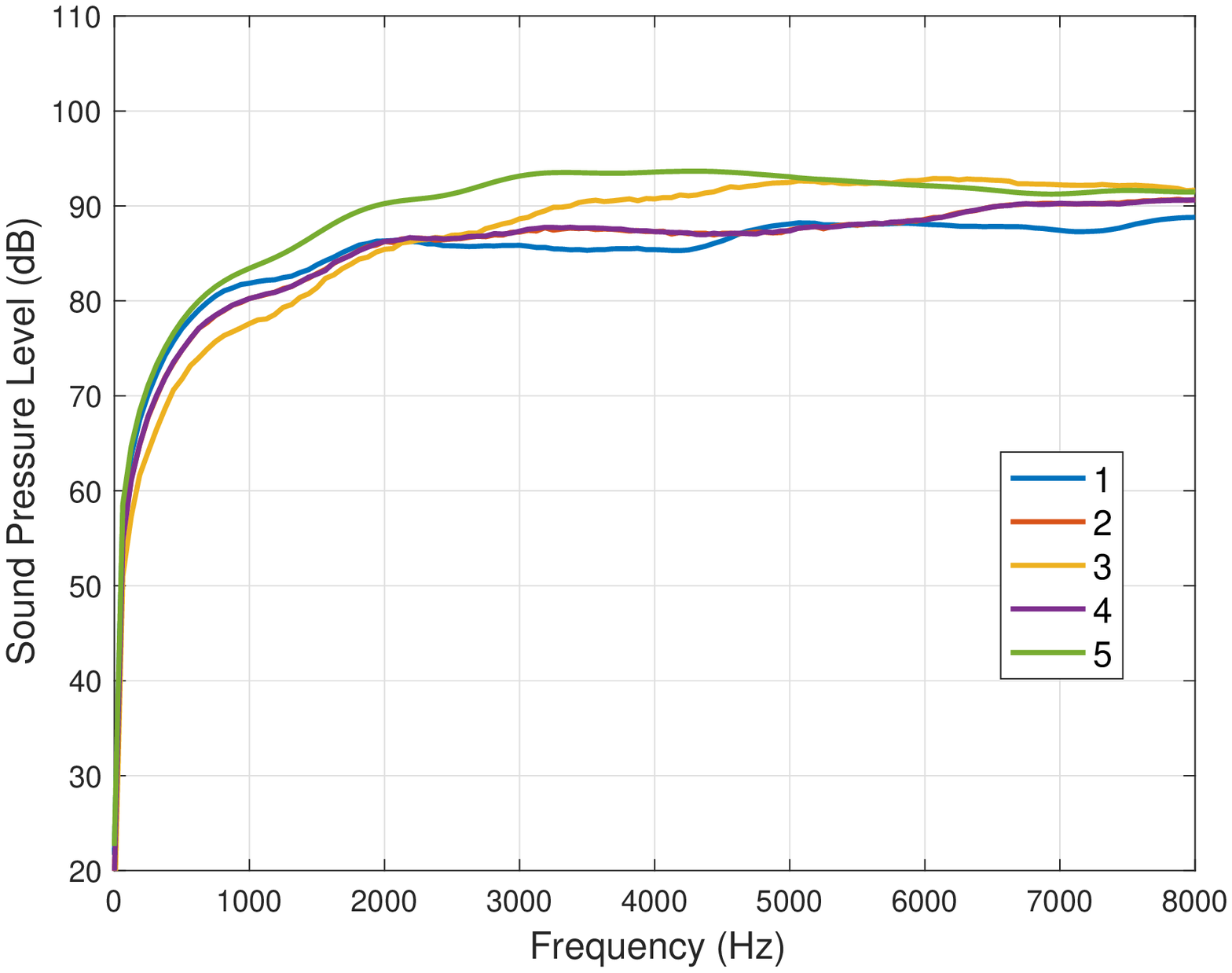} }}%
	\vspace*{-0.1cm}
    \caption{Magnitude of scattered wavefield, $p_s$, (in dB) at two plane wave angles.}
    \label{fig:scatterfield}%
\end{figure}

The big deviation of the FF performance at $\theta = 30^ \circ$ is attributed to the magnitude of the scattered wavefield (which is ignored in the FF case). This is illustrated in Fig. \ref{fig:scatterfield}, where we show the magnitude of the scattered wavefield at the five microphones at both angles (where the background plane wave has the same magnitude in both cases). Note that, the scattered wavefield at $\theta = 30^\circ$ is approximately $10$ dB stronger than $\theta = 90^\circ$ especially at high frequencies, which is manifested clearly in the corresponding AG/WNG/MACC behavior. This significant deviation of the free-field case demonstrates the limitation of this modeling and the necessity of incorporating the scattered field component through FEM modeling for beamformer design. 

\vspace{-0.3cm}

\section{Conclusion and Future Work}
The free-field model does not provide accurate modeling for broadband beamformer design, especially when the scattered wavefield is significant. Therefore, designing beamformer metrics based on free-field modeling results in suboptimal performance.  To mitigate this issue, we described a simulation-based framework for modeling the total wavefield, which is shown to noticeably improve the beamformer design. The model is universal for any device surface, and it could be used for both near-field and far-field modeling by computing the steering vectors of spherical and plane waves, respectively. Future work will utilize the results of this work to develop novel design techniques for broadband beamformer and generic form-factors that are based on this realistic microphone array modeling \cite{guangdongComsol2019}. Additionally, we expand the array processing metrics, and show a close matching of simulated and measured beampatterns for our proposed method \cite{guangdongComsol2019}.

\bibliographystyle{IEEEbib}
\bibliography{refs}

\begin{thebibliography}{10}

\bibitem{chhetri2018multichannel}
Amit Chhetri, Philip Hilmes, Trausti Kristjansson, Wai Chu, Mohamed Mansour,
  Xiaoxue Li, and Xianxian Zhang,
\newblock ``Multichannel {A}udio {F}ront-{E}nd for {F}ar-{F}ield {A}utomatic
  {S}peech {R}ecognition,''
\newblock in {\em 2018 26th European Signal Processing Conference (EUSIPCO)}.
  IEEE, 2018, pp. 1527--1531.

\bibitem{MA_book}
Michael Brandstein and Darren Ward,
\newblock {\em Microphone arrays: signal processing techniques and
  applications},
\newblock Springer Science \& Business Media, 2013.

\bibitem{benesty2008microphone}
Jacob Benesty, Jingdong Chen, and Yiteng Huang,
\newblock {\em Microphone array signal processing}, vol.~1,
\newblock Springer Science \& Business Media, 2008.

\bibitem{frost_FilterandSum}
Otis~Lamont Frost,
\newblock ``An algorithm for linearly constrained adaptive array processing,''
\newblock {\em Proceedings of the IEEE}, vol. 60, no. 8, pp. 926--935, 1972.

\bibitem{DS}
Dan~E Dudgeon,
\newblock ``Fundamentals of digital array processing,''
\newblock {\em Proceedings of the IEEE}, vol. 65, no. 6, pp. 898--904, 1977.

\bibitem{MVDR1}
Jack Capon,
\newblock ``High-resolution frequency-wavenumber spectrum analysis,''
\newblock {\em Proceedings of the IEEE}, vol. 57, no. 8, pp. 1408--1418, 1969.

\bibitem{MVDR2}
Henry Cox, Robert Zeskind, and Mark Owen,
\newblock ``Robust adaptive beamforming,''
\newblock {\em IEEE Transactions on Acoustics, Speech, and Signal Processing},
  vol. 35, no. 10, pp. 1365--1376, 1987.

\bibitem{RoomAcousticsBook}
Heinrich Kuttruff,
\newblock {\em Room acoustics},
\newblock {CRC} Press, fourth edition, 2000.

\bibitem{larsson2008partial}
Stig Larsson and Vidar Thom{\'e}e,
\newblock {\em Partial differential equations with numerical methods}, vol.~45,
\newblock Springer Science \& Business Media, 2008.

\bibitem{zotkin2017incident}
Dmitry~N Zotkin, Nail~A Gumerov, and Ramani Duraiswami,
\newblock ``Incident field recovery for an arbitrary-shaped scatterer,''
\newblock in {\em Acoustics, Speech and Signal Processing (ICASSP), 2017 IEEE
  International Conference on}. IEEE, 2017, pp. 451--455.

\bibitem{teutsch2007modal}
Heinz Teutsch,
\newblock {\em Modal array signal processing: principles and applications of
  acoustic wavefield decomposition}, vol. 348,
\newblock Springer, 2007.

\bibitem{rafaely2015fundamentals}
Boaz Rafaely,
\newblock {\em Fundamentals of spherical array processing}, vol.~8,
\newblock Springer, 2015.

\bibitem{VanTrees2002}
H.~L. {Van Trees},
\newblock {\em Optimum Array Processing},
\newblock Wiley, New York, 2002.

\bibitem{mabande2009design}
Edwin Mabande, Adrian Schad, and Walter Kellermann,
\newblock ``Design of robust superdirective beamformers as a convex
  optimization problem,''
\newblock in {\em Acoustics, Speech and Signal Processing, 2009. ICASSP 2009.
  IEEE International Conference on}. IEEE, 2009, pp. 77--80.

\bibitem{mabande2010design}
E~Mabande and Walter Kellermann,
\newblock ``Design of robust polynomial beamformers as a convex optimization
  problem,''
\newblock in {\em Proc. IEEE Int. Workshop Acoustic Echo, Noise Control
  (IWAENC)}, 2010, pp. 1--4.

\bibitem{acousticsbook}
Lawrence~E Kinsler, Austin~R Frey, Alan~B Coppens, and James~V Sanders,
\newblock {\em Fundamentals of acoustics},
\newblock Wiley, third edition, 1982.

\bibitem{FourierAcoustics}
Earl~G Williams,
\newblock {\em Fourier acoustics: sound radiation and nearfield acoustical
  holography},
\newblock Academic press, 1999.

\bibitem{mansour2018information}
Mohamed~F Mansour,
\newblock ``Information measures for microphone arrays,''
\newblock {\em arXiv preprint arXiv:1801.10128}, 2018.

\bibitem{pwd1}
Andrea Moiola, Ralf Hiptmair, and I~Perugia,
\newblock ``Plane wave approximation of homogeneous helmholtz solutions,''
\newblock {\em Zeitschrift f{\"u}r angewandte Mathematik und Physik}, vol. 62,
  no. 5, pp. 809, 2011.

\bibitem{pwd2}
Orhan Yilmaz and M~Turhan Taner,
\newblock ``Discrete plane-wave decomposition by least-mean-square-error
  method,''
\newblock {\em Geophysics}, vol. 59, no. 6, pp. 973--982, 1994.

\bibitem{pwd3}
Emmanuel Perrey-Debain,
\newblock ``Plane wave decomposition in the unit disc: Convergence estimates
  and computational aspects,''
\newblock {\em Journal of Computational and Applied Mathematics}, vol. 193, no.
  1, pp. 140--156, 2006.

\bibitem{berenger1994perfectly}
Jean-Pierre Berenger,
\newblock ``A perfectly matched layer for the absorption of electromagnetic
  waves,''
\newblock {\em Journal of computational physics}, vol. 114, no. 2, pp.
  185--200, 1994.

\bibitem{COMSOL}
COMSOL Multiphysics,
\newblock ``Acoustic module--user guide,'' 2017.

\bibitem{Bowman_book}
John~J Bowman, Thomas~B Senior, and Piergiorgio~L Uslenghi,
\newblock {\em Electromagnetic and acoustic scattering by simple shapes},
\newblock North-Holland Publishing Company, 1970.

\bibitem{wiener1949diffraction}
Francis~M Wiener,
\newblock ``The diffraction of sound by rigid disks and rigid square plates,''
\newblock {\em The Journal of the Acoustical Society of America}, vol. 21, no.
  4, pp. 334--347, 1949.

\bibitem{spence1948diffraction}
RD~Spence,
\newblock ``The diffraction of sound by circular disks and apertures,''
\newblock {\em The Journal of the Acoustical Society of America}, vol. 20, no.
  4, pp. 380--386, 1948.

\bibitem{guangdongComsol2019}
Guangdong Pan, Wontak Kim, Savaskan Bulek, Amit Chhetri, and Mohamed Mansour,
\newblock ``A study on acoustic modeling for microphone array beamforming,''
\newblock {\em pre-print}, 2019.

\end{thebibliography}

\end{document}